# Detecting Ambiguity Aversion in Cyberattack Behavior to Inform Cognitive Defense Strategies

Stephan Carney[1], Soham Hans[2], Sofia Hirschmann[3], Stacey Marsella[3], Yvonne Fonken[4], Peggy Wu[4], and Nikolos Gurney[2]

[1]Marshall School of Business, University of Southern California, Los Angeles, CA, USA
[2]Institute Creative Technologies, University of Southern California, Playa Vista, CA, USA
[3]Khoury College of Computer Sciences, Northeastern University, Boston, MA, USA
[4]Raytheon Technologies, USA

**ABSTRACT**

Adversaries (hackers) attempting to infiltrate networks frequently face uncertainty in their operational environments. This research explores the ability to model and detect when they exhibit ambiguity aversion, a cognitive bias reflecting a preference for known (versus unknown) probabilities. We introduce a novel methodological framework that (1) leverages rich, multi-modal data from human-subjects red-team experiments, (2) employs a large language model (LLM) pipeline to parse unstructured logs into MITRE ATT&CK-mapped action sequences, and (3) applies a new computational model to infer an attacker's ambiguity aversion level in near-real time. By operationalizing this cognitive trait, our work provides a foundational component for developing adaptive cognitive defense strategies.

## INTRODUCTION

Modern cybersecurity has evolved beyond a purely technological arms race into a complex, strategic interaction between human actors. While technical fortifications remain a necessary foundation for security, the inherent asymmetry of cyber conflict in which a defender must protect all assets while an attacker need only find one vulnerability necessitates a paradigm shift. A growing body of research suggests this asymmetry can be rebalanced by understanding, anticipating, and even exploiting the cognitive vulnerabilities and inherent biases of human attackers. Traditional security measures, which are often static and signature-based, prove insufficient against adversaries that employ novel and adaptive tactics. Consequently, a new frontier in cyber defense is emerging, one that models the adversary not as a monolithic, perfectly rational actor, but as a human decision-maker subject to the same psychological pressures and heuristics that govern behavior in all other domains.

## BEYOND RISK: AMBIGUITY IN ADVERSARIAL ENVIRONMENTS

A central challenge in modeling adversary cognition is accurately characterizing the environment in which they operate. Much of the existing work implicitly frames the attacker's problem as one of decision-making under *risk*, where the probabilities of various outcomes (e.g., an exploit succeeding, an action being detected) are known or can be reasonably estimated. However, this paper contends that the operational reality for an attacker is far better described by the concept of *ambiguity*, a state of uncertainty where such probabilities are unknown, incalculable, or unreliable (Knight, 1921).

Ambiguity (or Knightian uncertainty) arises in situations that are novel, complex, and/or lack sufficient historical data to form a stable probability distribution over outcomes. An attacker navigating an unfamiliar network, encountering bespoke security configurations or predicting the response of a human network administrator faces precisely this kind of uncertainty. Their decisions are less like betting on a fair coin flip (risk), but rather akin to betting on an unfair dice roll with unknown and unknowable bias (ambiguity). This distinction is not merely semantic; it points to fundamentally different psychological mechanisms governing choice and the downstream modelling decisions.

## Ambiguity Aversion versus Loss Aversion

Current research in cognitive cybersecurity aims to exploit cognitive biases or vulnerabilities for psychology-informed defense. A current focus of research like the IARPA ReSCIND program has been on loss aversion, a well-documented bias where individuals weigh potential losses more heavily than equivalent gains (Hans et al., 2025). Models grounded in prospect theory (Kahneman & Tversky, 1979) operationalize loss aversion by having an agent evaluate the subjective utility of actions based on estimated gains, losses, and their associated probabilities (Huang et al., 2024). For example, the probability of an attacker choosing an aggressive action, $\gamma$, is modeled as a function of the subjective utility, $\epsilon$, of aggressive ($a$) versus stealty ($s$) options, where ($e$) itself is a function of a loss aversion coefficient, $\lambda_l$:

$$\gamma(a, s, \lambda_l) \coloneqq \frac{1}{1 + e^{-\mu(\epsilon(a,\lambda_l) - \epsilon(s,\lambda_l))}}$$

While valuable, such models presuppose an environment of *risk*, where the parameters for gain, loss, and probability are accessible to the agent's decision calculus.

This work proposes a shift in focus to ambiguity aversion, defined as the preference for options with known probabilities over those with unknown probabilities. This phenomenon is famously illustrated by the Ellsberg Paradox (Ellsberg, 1961). In its classic formulation, individuals are asked to bet on drawing a colored ball from one of two urns. Urn A contains 50 red and 50 black balls (a known risk). Urn B contains 100 red and black balls in an unknown proportion (an ambiguous decision). A robust finding is that people will systematically prefer to bet on a color from Urn A, regardless of which color they choose, thereby avoiding the ambiguity of Urn B.

This preference for the known over the unknown is profoundly relevant to cybersecurity. An attacker's choice between exploiting a well-documented vulnerability in a standard Windows server versus probing a custom-built, undocumented application is a direct analogue of the Ellsberg choice. The former represents a known risk, while the latter is fraught with ambiguity. Therefore, we argue that ambiguity aversion is a more fundamental and ecologically valid construct for modeling attacker behavior in the wild. It captures the core psychological challenge of acting in the face of irreducible uncertainty, a challenge that is often more pressing than the simple weighing of potential gains and losses. Table 1 provides a clear conceptual distinction between these related but distinct cognitive phenomena.

**Table 1.** Risk, Loss, and Ambiguity Aversion

| Concept | Psychological Mechanism | Cybersecurity Example |
| --- | --- | --- |
| **Risk Aversion** | Preference for a certain outcome over a gamble with equal or higher expected value. | Choosing a reliable exploit with a 90% success rate for moderate gain over a novel exploit with a 50% success rate for a much higher gain. |
| **Loss Aversion** | Weighing potential losses more heavily than equivalent gains. | Refusing to use a noisy but effective exploit that risks revealing an established foothold (losing access), even if the potential gain is significant. |
| **Ambiguity Aversion** | Preference for an option with known probabilities over one with unknown probabilities. | Choosing to attack a standard, well-documented software service over a custom, in-house application with unknown security configurations. |

## METHODOLOGY

Our methodology provides an end-to-end pipeline for inferring the latent cognitive trait of ambiguity aversion from raw, human-generated behavioral data. This process is composed of three distinct but interconnected stages: (1) leveraging a high-fidelity experimental dataset as a foundation, (2) using a large language model to translate unstructured data into a structured format, and (3) applying a theory-driven computational model to infer the cognitive bias. This modular architecture represents a generalizable template for a new class of cognitive security sensors.

## The GAMBiT Experiments

The empirical grounding for our methodology is the rich, multi-modal dataset generated by the Guarding Against Malicious Biased Threats (GAMBiT) project (Beltz et al., 2025). These experiments were designed to capture attacker decision-making with high ecological validity. The experimental setup involved a series of human-subjects red-team exercises conducted in a realistic enterprise cyber range provisioned on the SimSpace Cyber Force Platform. Each participant was placed in an isolated network environment comprising approximately 40 virtual hosts populated with representative business services and background user traffic.

Over the course of two 8-hour sessions, participants (N=19-20 per experiment) were tasked with pursuing broad operational objectives, such as network infiltration, privilege escalation, and data exfiltration. A critical data modality for our work is the collection of detailed, timestamped *operation notes*. In these free-text logs, participants documented their thought processes, tactical decisions, tool usage, and justifications for their actions, providing a window into their real-time decision-making calculus.

Crucially, the GAMBIT experimental design incorporated both control and experimental conditions. Experiment 2 served as a control, with participants operating in a baseline environment. Experiments 1 and 3 introduced a suite of "cognitive triggers:" deceptive artifacts, misleading files, and risky shortcuts designed to elicit and measure specific biases, including loss aversion and confirmation bias. This experimental structure provides a valuable pathway for the validation of our ambiguity aversion model against ground-truth behavioral data linked to specific cognitive stimuli.

**LLM-Powered Behavioral Annotation**

The second stage of our pipeline addresses the fundamental challenge of transforming the raw, unstructured *operation notes* (i.e., OpNotes) and NetFlow log files (via Suricata) into a structured time-series of attacker actions suitable for computational analysis. OpNotes are simply notes kept by study participants of their efforts. These are similar in nature to documented note taking of attackers. Suricata is a powerful, open-source Network Intrusion Detection and Prevention System (NIDS/NIPS) that uses deep packet inspection to analyze network traffic, identify and block threats, log activities, and capture files (OISF, 2025). This process is based on the LLM-based annotation pipeline developed by Hans et al. (2025) and used in other annotation pipelines (Kim et al., 2025). It leverages the advanced reasoning capabilities of large language models to act as a universal translator for human behavioral data.

**PsychSim Framework and the Computational Model of Ambiguity Aversion**

The final stage of the pipeline is a specialized computational model implemented within the Gambit PsychSim framework, a multi-agent system designed to infer and exploit cognitive biases. This framework models two primary agents: an Attacker and a Defender. The Defender's objective is to form a mental model of the Attacker by observing their actions and using those observations to infer cognitive biases that can be subsequently exploited. The underlying agent technology in PsychSim is a form of Partially Observable Markov Decision Process (POMDP), which provides agents with an observation function, a reward function, and the capacity for sequential decision-making to maximize expected utility (Marsella et al., 2004; Pynadath and Marsella, 2005). Critically, PsychSim enables Theory of Mind (ToM) reasoning through recursive modeling, allowing the Defender agent to possess and update embedded models of the Attacker.

As part of the GAMBiT project, the PsychSim agent has been used to fit five models of cognitive biases: sunk cost fallacy, base rate neglect, confirmation bias, the availability heuristic, and loss aversion (Hirschmann et al., 2025). It models sunk cost fallacy by updating based on repeated target engagement and estimated future value. The implementation considers host value, difficulty, and investment already made (Kleinberg et al., 2021). Base rate neglect is treated as the ignoring or under-weighting the prior probability (base rate) and over-weighing new evidence (Bar-Hillel, 1980). The agent observes actions aligned with confirmation bias when the attacker persists with an exploit despite accumulating evidence of its poor success rate relative to their initial expectations (Wason, 1960). To model the availability heuristic, the agent uses a lexical analysis that seeks to assess whether the attacker targets account names that suggest administrative privileges, with higher scores for "admin" or "room" compared to personal names. This assumes attackers' perceptions are affected by what is familiar and expected (Yuill, Denning, and Feer, 2007). Finally, the agent takes a simplified approach and models loss aversion as preferring less risky actions, such as avoiding actions that have a high risk of discovery (Ert an Erev, 2013).

Our computational model for ambiguity aversion serves as a heuristic update rule within this observation-driven process. The model's core logic is grounded in the comparative ignorance hypothesis (Fox and Tversky, 1995), which posits that ambiguity aversion is most pronounced when a decision-maker must directly compare a vague prospect with a clearer one. Accordingly, our model activates its inferential logic primarily in high-uncertainty scenarios where the attacker faces a choice between a familiar, safer action and a less familiar, ambiguous alternative. The contrast between these states of knowledge is what reveals the underlying bias.

Following the comparative foundation for ambiguity aversion proposed by Ghirardato and Marinacci (2002), our model operationalizes this bias by treating Subjective Expected Utility (SEU) as the benchmark for ambiguity-neutral behavior. The model's score represents a quantifiable measure of the attacker's deviation from this SEU baseline. An action is classified as ambiguity-averse if the attacker, when faced with significant uncertainty, chooses a safer option over a potentially higher-reward but more ambiguous alternative—a pattern inconsistent with SEU maximization.

To quantify this crucial element of uncertainty, the model employs a set of behavioral heuristics. Rather than assuming a known probability distribution, as the current instantiation of the loss aversion model used in the GAMBiT project, it estimates the degree of ambiguity by considering factors that reflect the attacker's relative ignorance or lack of competence regarding an action. These factors include the novelty of the action or target, the inherent complexity of the technique being employed, and the historical variance in the action's success rate. A high score on these dimensions suggests a state of Knightian uncertainty, where the attacker cannot reliably assess the probability of success. The model also identifies more sophisticated behavioral signatures, such as hedging as an indicator of an attempt to manage ambiguity. By grounding its heuristics in established theories of decision-making under ambiguity, the model provides a transparent and interpretable method for inferring a latent cognitive trait from observable behavior.

## RESULTS

To evaluate our ambiguity aversion model, we ran it on the behavioral data from the GAMBIT experiments, generating probability estimates for each participant's actions (1583 observations from 29 participants). We then conducted an exploratory analysis comparing these estimates to those produced by the established loss aversion model developed within the GAMBiT project (Beltz et al., 2025).

First, we compared the overall distribution of trait probabilities generated by each model. As shown in Table 2 and the density plot in Figure 1, the loss aversion model tended to assign higher probabilities across all observations. A Wilcoxon signed-rank test confirmed this visual finding, indicating that the probabilities from the loss aversion model were significantly higher than those from the ambiguity aversion model (W = 450097, $p < .001$).

**Table 2.** Trait Probability by Model

| Model | Mean (SD) Trait Probability | Median Trait Probability | Percentage of Probabilities > 0.5 |
|---|---|---|---|
| **Ambiguity Aversion** | 0.212 (0.196) | 0.091 | 15.0% |
| **Loss Aversion** | 0.439 (0.075) | 0.478 | 0.0% |

Next, we examined the models' ability to generate high-confidence signals. While the loss aversion model produced higher mean probabilities, the ambiguity aversion model generated a greater number of high-confidence observations, defined as a probability greater than 0.5. The ambiguity aversion model yielded 237 high-confidence observations, compared to 0 from the loss aversion model (14.97% vs. 0%; $\chi^2(0) = 254.02$, $p < .001$). This suggests that while the loss aversion model is more broadly applied, our ambiguity aversion model provides more decisive and confident estimates when it detects the relevant behavioral signals.

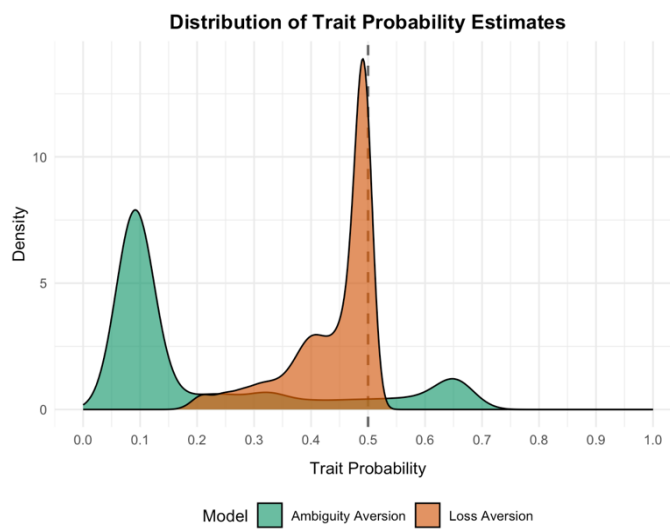

**Figure 1.** Density plot displaying the distribution of trait probability estimates by model.

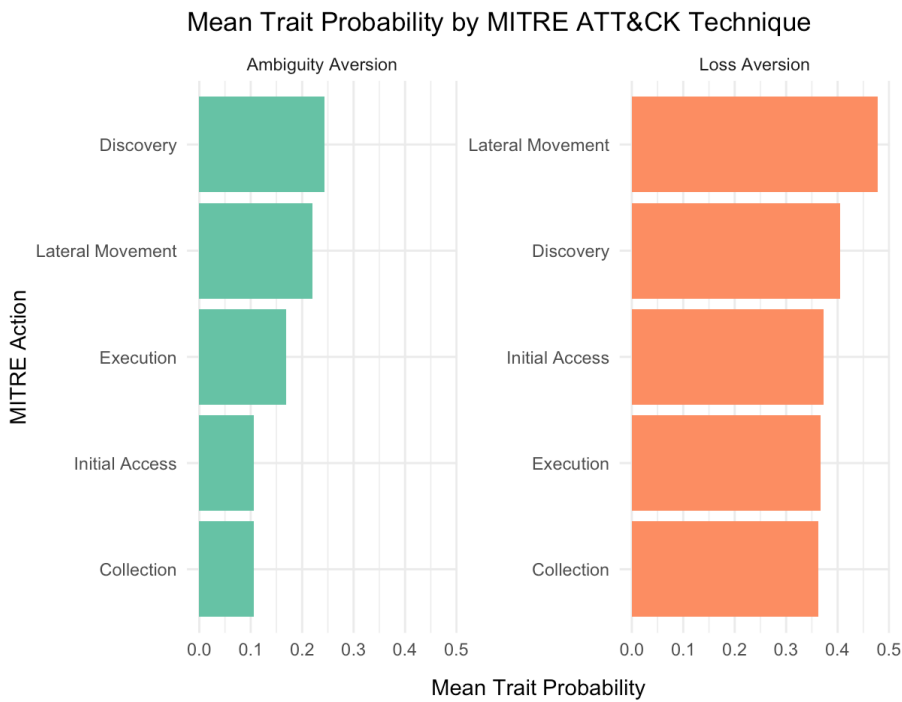

**Figure 2.** Mean trait probability by MITRE ATT&CK Technique across each model.

Finally, we analyzed how the inferred trait probabilities varied across different MITRE ATT&CK tactics (see Figure 2). The loss aversion model assigned the highest average probabilities to actions associated with Lateral Movement tactics. In contrast, the ambiguity aversion model's highest probabilities were linked to Discovery tactics. This divergence suggests that the two models are capturing distinct behavioral drivers that manifest at different stages of a cyberattack, with loss aversion being more prominent during evasion, and ambiguity aversion being more influential during initial reconnaissance and action execution.

## DISCUSSION

The initial finding that loss aversion explains more behavioral variance than ambiguity aversion is, we believe, a direct consequence of the experimental environment's design. Our central hypothesis is that the simulated cyber range, while ecologically valid in many respects, did not instantiate a sufficiently high risk associated with ambiguity. In the context of the GAMBIT experiments, the negative outcomes of choosing an ambiguous, high-uncertainty path—such as a failed exploit or time spent on a dead end—lacked significant weight. There was no tangible penalty equivalent to "getting caught" or losing a persistent foothold, which are the primary drivers that would make an attacker ambiguity-averse. Consequently, the more immediate and salient concern for participants was likely the preservation of existing access, a behavior directly captured by loss aversion models. This is particularly true of the GAMBiT PsychSim model, which primarily models risk aversion.

It is also critical to acknowledge the preliminary nature of this work. The cognitive models for both ambiguity and loss aversion require substantial training on larger datasets to be properly calibrated. The current parameter space for our ambiguity aversion model was largely exploratory, established without strong guiding principles from prior empirical data. With more data, these values can be systematically set and validated.

Furthermore, true validation of these cognitive models will require a dynamic, interactive environment. A definitive test of whether the parameterization is correct can only occur when a defensive agent, guided by the model's inferences, attempts to actively manipulate an attacker's biases and observes the resulting behavioral change.

## CONCLUSION AND FUTURE WORK

This paper has presented a novel, complete, and empirically grounded methodology for detecting and quantifying ambiguity aversion in cyberattacker behavior. We have advanced the argument that, within the context of sophisticated cyber operations characterized by Knightian uncertainty, ambiguity aversion is a theoretically sound and ecologically valid construct. We detailed a three-stage pipeline that transforms raw, unstructured human behavioral data into a real-time inference about a latent cognitive trait. While our preliminary analysis shows that a standard loss aversion model currently offers greater explanatory power, we posit this is an artifact of the experimental conditions rather than a fundamental refutation of ambiguity aversion's role in cyber operations.

### Future Work

This research opens several promising avenues for future work:
- **Experiment Design for Ambiguity:** The most critical direction for future work is the design and execution of a new human-subjects experiment where the risk of ambiguity is made explicit and consequential. This experiment must incorporate a tangible risk of "getting caught," such as a high probability of being detected and ejected from the network upon failing an exploit on an ambiguous, unknown system. This will create the necessary conditions to properly elicit and measure ambiguity-averse behavior.
- **Model Training and Calibration:** Both the ambiguity and loss aversion models must be formally trained and validated. This involves collecting more extensive behavioral data to establish robust guiding principles for the parameter space, moving beyond the current exploratory settings.
- **Interactive Validation:** The ultimate validation of this framework lies in its application. Future work should focus on integrating these cognitive sensors into an interactive defensive agent. This agent would use the inferred biases to deploy cognitive triggers and measure the subsequent change in attacker behavior, thereby closing the loop and confirming the model's predictive and manipulative utility.

**ACKNOWLEDGMENT**

This research is based upon work supported in part by the Office of the Director of National Intelligence (ODNI), Intelligence Advanced Research Projects Activity (IARPA) under Reimagining Security with Cyberpsychology Informed Network Defenses (ReSCIND) program contract N66001-24-C-4504. The views and conclusions contained herein are those of the authors and should not be interpreted as necessarily representing the official policies, either expressed or implied, of ODNI, IARPA, or the U.S. Government. The U.S. Government is authorized to reproduce and distribute reprints for governmental purposes notwithstanding any copyright annotation therein.